\documentclass[twocolumn,showpacs,amssymb]{revtex4}

\usepackage{latexsym}

\begin{document}

\title{$\zeta$-function regularization and the interacting
Bose gas at low temperature}

\author{David J. Toms}
  \email{d.j.toms@newcastle.ac.uk}
\affiliation{Department of Physics, University of Newcastle Upon Tyne,\\
Newcastle Upon Tyne, United Kingdom NE1 7RU}

\date{\today}

\begin{abstract}
$\zeta$-function methods are used to study the properties of the
non-relativistic interacting Bose gas at finite temperature and
density. Results for the ground state energy and pressure are
obtained at both zero and finite temperature. The method used does
not restrict the form of the interaction, which can be completely
general. A similar procedure is then applied to evaluate the
ground state energy of a binary mixture consisting of different
bosons. Analytical results are obtained in the case where the two
species have the same mass.
\end{abstract}

 \maketitle

\def\bea{\begin{eqnarray}}
\def\eea{\end{eqnarray}}
\def\be{\begin{equation}}
\def\ee{\end{equation}}
\def\nn{\nonumber\\}
\def\bp{\bar{\Psi}}
\def\b2{|\bp|^2}
\def\x{\mathbf{x}}

\section{Introduction}
\label{sec1}

The non-interacting Bose gas, and its behaviour as a function of
temperature, is a key part of any study of statistical mechanics.
(See for example, Ref.~\cite{London,LLSP,Huang,Pathria}.)  However
the neglect of interactions among the particles, which may be
justified as a first approximation, cannot be valid at any
fundamental level.  (As emphasised in Ref.~\cite{Huang}, the
condensation of a free Bose gas has unphysical properties, such as
infinite compressibility.)  Accordingly the study of interparticle
interactions is of obvious physical interest.

The interacting Bose gas received a great deal of attention about
forty years ago.  Textbook treatments may now be found.  (See for
example Ref.~\cite{Huang,Pathria,FetterWalecka}.)  The methods
used to study the interacting Bose gas go beyond what would
normally be described as elementary.  The pioneering studies
include \cite{LeeYang,BruecknerSawada,LeeHuangYang} who obtained
the ground state energy of the interacting Bose gas at zero
temperature using a variety of methods.  The generalization to
$T\ne0$ was performed by Lee and Yang ~\cite{LeeYangII}.  Much
more recently it was shown how modern functional integral methods
of quantum field theory could be used to study the interacting
Bose gas to great effect~\cite{HHR}. It was also shown \cite{BBD}
how the results for the non-relativistic interacting Bose gas
could be obtained as a limiting case of the relativistic gas.

The main purpose of the present paper is to show how it is
possible to study the interacting Bose gas at low temperatures and
densities in a relatively straightforward way.  The method makes
use of the background field method~\cite{DeWitt} to obtain the
excitation spectrum of the gas in a manner akin to that of
Bernstein and Dodelson~\cite{BernsteinDodelson}.  There is no need
to choose a specific form for the interaction.  The zero
temperature result for the ground state energy can then be
understood as a regularized zero-point energy and studied in a
similar way to the Casimir effect~\cite{Casimir}.  By adopting
$\zeta $-function regularization~\cite{DowkerCritchley,Hawking}
and specialising the form of the interaction, is easy to derive
the classic result \cite{LeeYang,BruecknerSawada,LeeHuangYang} for
the ground state energy.  Such a technique renders the discussion
of renormalization of the theory trivial.  By an examination of
the thermodynamic potential it is possible to obtain finite
temperature corrections to the zero temperature results in a
straightforward way.

To further exemplify the methods used we will also study a mixture
of two interacting Bose gases.  Although there has been some work
on the subject \cite{Bassichis,Nepo,Colson} so far as we are
aware, the analogue of the ground state energy calculations of
Ref.~\cite{LeeYang,BruecknerSawada,LeeHuangYang} have not been
given before.  We do this calculation for $T = 0$ as well as
$T\ne0$ below.  It is worth pointing out that mixtures of two Bose
gases have been achieved experimentally for trapped atomic gases
\cite{Myatt}. (A small sample of some of the theoretical work on
trapped mixtures includes \cite{HoShenoy,Esry,Pu,Ao}.)

\section{Single species gas}
\label{sec2}
\subsection{General approach}
\label{sec2.1}

In order to see how the method of $\zeta$-function regularization
works in a simple setting we will first study the interacting
single species gas. In the case where the interaction is quartic
in the fields it will be shown how the classic low temperature
results of
Refs.~\cite{LeeYang,BruecknerSawada,LeeHuangYang,LeeYangII} are
found in a relatively easy manner. After describing the basic
setup in this section, we will obtain the ground state energy and
the equation of state in Secs.~\ref{sec2.2} and \ref{sec2.3}.

We begin with the action functional for a single complex scalar
field $\Psi$ \bea S\lbrack\Psi,\Psi^\dagger\rbrack&=&\int\,dt
\int_{\Sigma}d\sigma_x
\Big\lbrace\frac{i}{2}(\Psi^\dagger\dot{\Psi}-\dot{\Psi}^\dagger\Psi)\nn
&&-\frac{1}{2m}|\nabla\Psi|^2
+\mu|\Psi|^2-U(|\Psi|^2)\Big\rbrace\;.\label{2.1} \eea Here
$\Sigma$ denotes the space of interest. Our concern here will be
with flat 3-dimensional space, so that $\Sigma$ can be thought of
as a finite dimensional box with the infinite volume limit taken.
The chemical potential $\mu$ enforces the conserved particle
number. $U(|\Psi|^2)$ is the interaction potential which we allow
to be an arbitrary function. As the notation suggests, we assume
that $U$ depends on $\Psi$ and $\Psi^\dagger$ only through
$|\Psi|^2$, and we take $U$ to be real.

The thermodynamic potential $\Omega$ at lowest order in  quantum
field theory is comprised of three pieces~: \be \Omega=\Omega_{\rm
class}+\Omega^{(1)}_{T=0}+ \Omega^{(1)}_{T\ne0} \;.\label{2.2} \ee
Here \be \Omega_{\rm class}=V\left(U(\b2)-\mu\b2\right)\label{2.3}
\ee is the contribution from the classical part of the theory with
$\bp$ representing any non-zero condensate. ($\bp$ can be thought
of as the background field in the background field
method~\cite{DeWitt}.) We take $\bp$ to be constant here. $V$ is
the volume of $\Sigma$. The terms in $\Omega^{(1)}_{T=0}$ and
$\Omega^{(1)}_{T\ne0} $ are the first order quantum corrections to
the classical theory, with $\Omega^{(1)}_{T\ne0}$ representing the
finite temperature corrections to the zero temperature theory
described by $\Omega^{(1)}_{T=0}$. The last two terms of
(\ref{2.2}) can be obtained in a consistent development using the
background-field method; however, we will use a more elementary
approach here.

We will use the method of excitation energies as described  in
Ref.~\cite{BernsteinDodelson}. Begin with the equation of motion
for $\Psi$ and $\Psi^\dagger$ which follow from (\ref{2.1}) as
\bea
0&=&i\dot{\Psi}+\frac{1}{2m}\nabla^2\Psi+ \mu\Psi- U'(|\Psi|^2)\Psi\;,\label{2.4a}\\
0&=&-i\dot{\Psi}^\dagger+\frac{1}{2m}\nabla^2\Psi^\dagger+
\mu\Psi^\dagger-U'(|\Psi|^2)\Psi^\dagger\;.\label{2.4b} \eea If we
expand $\Psi=\bp+\Phi$, where $\Phi$ represents the fluctuation
about the background field $\bp$, then $\Phi$ satisfies \bea
0&=&i\dot{\Phi}+\frac{1}{2m}\nabla^2\Phi+ \mu\Phi-U'(\b2)\Phi\nn
&&\quad-U''(\b2)\left(\b2\Phi+\bp^2\Phi^\dagger\right)\;,\label{2.5a}\\
0&=&-i\dot{\Phi}^\dagger+\frac{1}{2m}\nabla^2\Phi^\dagger+
\mu\Phi^\dagger-U'(\b2)\Phi^\dagger\nn
&&\quad-U''(\b2)\left(\b2\Phi^\dagger+(\bp^\dagger)^2\Phi
\right)\;.\label{2.5b} \eea If we define $\lbrace f_k(\x)\rbrace$
to be a complete orthonormal  set of solutions to \be
-\nabla^2f_k(\x)=\sigma_kf_k(\x)\;,\label{2.6} \ee we can expand
\bea
\Phi(t,\x)&=&\sum_{k}A_ke^{-iE_kt}f_k(\x)\;,\label{2.7a}\\
\Phi^\dagger(t,\x)&=&\sum_{k}B_ke^{-iE_kt}f_k(\x)\;,\label{2.7b}
\eea for independent expansion coefficients $A_k$ and $B_k$. $E_k$
are  the excitation energies. Because of the assumption of
completeness for the $f_k(\x)$ we find, after substitution of
(\ref{2.7a}) and (\ref{2.7b}) into (\ref{2.5a}) and (\ref{2.5b}),
\bea 0&=&\Big(E_k-\frac{\sigma_k}{2m}+\mu-U'(\b2)-\b2
U''(\b2)\Big)A_k\nn
&&\quad\quad -\bp^2 U''(\b2)B_k\;,\label{2.8a}\\
0&=&\Big(-E_k-\frac{\sigma_k}{2m}+\mu-U'(\b2)-\b2
U''(\b2)\Big)B_k\nn  &&\quad\quad-(\bp^{\dagger})^2
U''(\b2)A_k\;.\label{2.8b} \eea For a non-trivial solution for the
expansion coefficients $A_k$ and $B_k$ we must have \bea
E_k&=&\Big(\frac{\sigma_k}{2m}-\mu+U'(\b2)\Big)^{1/2}\label{2.9}\\
&&\times\Big(\frac{\sigma_k}{2m}-\mu+U'(\b2)+2\b2
U''(\b2)\Big)^{1/2} \;.\nonumber \eea

We can now write down the last two contributions to the
thermodynamic  potential (\ref{2.1}). We have \be
\Omega_{T=0}^{(1)}=\frac{\hbar}{2}\sum_kE_k\;,\label{2.10} \ee
since at zero temperature the only contribution is the sum over
zero-point energies. Also \be \Omega_{T\ne0}^{(1)}=\hbar
T\sum_k\ln(1-e^{-\beta E_k})\;,\label{2.11} \ee which is the
expression usually written down for the thermodynamic potential at
finite temperature. We leave the factor of $\hbar$ in front of
both expressions (\ref{2.10}) and (\ref{2.11}) to denote that they
are of first order in a systematic loop expansion. We will only
work to order $\hbar$.

$\Omega_{T=0}^{(1)}$ is divergent, and to make sense of this we
must regularize it in some way. We will use a version of
$\zeta$-function regularization \cite{DowkerCritchley,Hawking}
here. (See Ref.~\cite{DJTKorea} for a pedagogical introduction to
the method we will use.) This consists of defining an energy
$\zeta$-function by \be E(s)=\sum_kE_k(\ell
E_k)^{-s}\;,\label{2.12} \ee where $s$ is a complex variable
suitably chosen so that the sum in (\ref{2.12}) converges, and
$\ell$ is a constant with dimensions of inverse energy introduced
so that $E(s)$ has units of energy for any value of $s$. The
regularization consists of defining $\Omega_{T=0}^{(1)}$ by \be
\Omega_{T=0}^{(1)}=\frac{\hbar}{2}E(0)\;,\label{2.13} \ee where
$E(0)$ represents the analytic continuation of $E(s)$ from the
region of the complex $s$-plane where (\ref{2.12}) converges to
$s=0$, assuming that $E(s)$ is regular at $s=0$. It may be that
$E(s)$ has a simple pole at $s=0$ in which case $E(0)$ in
(\ref{2.13}) represents the analytic continuation of $E(s)$ to a
neighbourhood of $s=0$. The pole can be dealt with by the usual
machinery of renormalization. In the theory considered here
renormalization is not necessary as we will show that $E(s)$ is
analytic at $s=0$.

Given the thermodynamic potential $\Omega$, the number of
particles is given by \be
N=-\left.\left(\frac{\partial\Omega}{\partial\mu}\right)\right|_{\beta,V,\bp}
\;,\label{2.14} \ee and the internal energy density $\rho$ is \be
\rho V=\left.\frac{\partial}{\partial\beta}
(\beta\Omega)\right|_{\beta\mu,V,\bp} \;.\label{2.15} \ee Using
(\ref{2.14}) it is easily seen that \be \rho
V=\left.\frac{\partial}{\partial\beta}
(\beta\Omega)\right|_{\mu,V,\bp}+\mu N \;.\label{2.16} \ee In
addition, the condensate $\bp$ satisfies \be
0=\left.\frac{\partial\Omega}{\partial\bp}\right|_{\beta,V,\mu}
\;.\label{2.17} \ee The exact equation for $\bp$ is very
complicated due to the  dependence of the excitation energies
$E_k$ on $\bp$ in (\ref{2.9}). However because we are only working
to order $\hbar$, from (\ref{2.2}) and (\ref{2.3}) we can conclude
that \be \mu=U'(\b2)+{\mathcal O}(\hbar)\;.\label{2.18} \ee This
means that in any expression which is already of order $\hbar$ we
can simplify it using only the first term of (\ref{2.18}). Of
course in any term which is of lower order in $\hbar$ we must
include the ${\mathcal O}(\hbar)$ correction to $\mu$ for
consistency. We will encounter this in Sec.~\ref{sec2.3}.

\subsection{Ground state energy}
\label{sec2.2}

We will introduce the number density $n$ by \be
n=\frac{N}{V}\;.\label{2.19} \ee From
(\ref{2.2},\ref{2.3},\ref{2.11},\ref{2.14}) we find the  number
density of particles to be \be
n=\b2+n_{T=0}^{(1)}+n_{T\ne0}^{(1)}\;,\label{2.20} \ee with \be
n_{T=0}^{(1)}=-\frac{1}{V}\left.\left(
\frac{\partial\Omega_{T=0}^{(1)}}{\partial\mu}\right)\right|_{\beta,V,\bp}
\label{2.21} \ee and a similar expression holding for
$n_{T\ne0}^{(1)}$. Since $n_{T=0}^{(1)}$ and $n_{T\ne0}^{(1)}$ are
of order $\hbar$, it is obvious from (\ref{2.20}) that \be
\b2=n+{\mathcal O}(\hbar)\;.\label{2.23} \ee Using the same
argument as for $\mu$ below (\ref{2.18}), we can set $\b2=n$ in
any expression which is already of order $\hbar$.

If we use (\ref{2.2},\ref{2.3}) along with (\ref{2.19}) and
(\ref{2.20}) in the expression (\ref{2.16}) for the energy
density, we have \bea \rho&=&U(\b2)+\frac{1}{V}\Omega_{T=0}^{(1)}+
\frac{1}{V}\left.\frac{\partial}{\partial\beta}(\beta\Omega_{T\ne0}^{(1)})
\right|_{\mu,V,\bp}\nn
&&\quad\quad+\mu(n_{T=0}^{(1)}+n_{T\ne0}^{(1)})\;.\label{2.25}
\eea We can use (\ref{2.20}) to simplify the argument of the
energy $U(\b2)$. To order $\hbar$ we have \be
U(\b2)=U(n)-U'(n)(n_{T=0}^{(1)}+n_{T\ne0}^{(1)})+\cdots\;.\label{2.26}
\ee The last term of (\ref{2.25}) may be simplified by using
(\ref{2.18}) and (\ref{2.23}) to give \be
\mu(n_{T=0}^{(1)}+n_{T\ne0}^{(1)})=U'(n)(n_{T=0}^{(1)}+n_{T\ne0}^{(1)})
+\cdots\;,\label{2.27} \ee again working only to order $\hbar$. We
therefore find \be \rho=U(n)+\frac{1}{V}\Omega_{T=0}^{(1)}+
\left.\frac{1}{V}\frac{\partial}{\partial\beta}
(\beta\Omega_{T\ne0}^{1)})\right|_{\mu,V,\bp}\;,\label{2.28} \ee
from (\ref{2.25}), holding to order $\hbar$. If we use
(\ref{2.11}), then the last term of (\ref{2.28}) becomes \bea
\rho_{T\ne0}^{(1)}&=&\left.\frac{1}{V}
\frac{\partial}{\partial\beta}
(\beta\Omega_{T\ne0}^{(1)})\right|_{\mu,V,\bp}\nn\\
&=&\frac{\hbar}{V}\sum_{k}E_k(e^{\beta
E_k}-1)^{-1}\;.\label{2.28a} \eea This is the usual expression for
the internal energy density  from statistical mechanics.

We now require the evaluation of $\Omega_{T=0}^{(1)}$.
Eq.~(\ref{2.12})  can be used for this. A direct evaluation using
the energy levels in (\ref{2.9}) is difficult. Because we are only
working to order $\hbar$, and $\Omega_{T=0}^{(1)}$ is already of
order $\hbar$, we can simplify the excitation energies using
$\mu=U'(n)$ and $\b2=n$. This results in \be
E_k=\left(\frac{\sigma_k}{2m}\right)^{1/2}
\left(\frac{\sigma_k}{2m}+ 2nU''(n)\right)^{1/2}\;.\label{2.29}
\ee If we take the large volume limit we can replace
$\sigma_k\rightarrow k^2$ and $\sum_{k}\rightarrow
V\int\frac{d^3k}{(2\pi)^3}$; thus the energy $\zeta$-function to
be evaluated is \be
E(s)=\frac{V}{\ell^{s}}\int\frac{d^3k}{(2\pi)^3}
\left(\frac{k^2}{2m}\right)^{\frac{(1-s)}{2}}
\!\!\left(\frac{k^2}{2m}+
2nU''(n)\right)^{\frac{(1-s)}{2}}\;.\label{2.30} \ee (Note that
the energy spectrum in Eq.~(\ref{2.29}) with $\sigma_k\rightarrow
k^2$ was first given by Bogoliubov~\cite{Bogoliubov}.) The
integral in (\ref{2.30}) is easily evaluated using the
representation of the $\Gamma$-function to give \be
E(s)=(2m)^{3/2}\frac{V}{4\pi^2\ell^s}
\frac{\Gamma(2-\frac{s}{2})\Gamma(s-\frac{5}{2})}{\Gamma(\frac{s-1}{2})}
\lbrack2nU''(n)\rbrack^{5/2-s}\;.\label{2.31} \ee This expression
is analytic at $s=0$ with the result \be
E(0)=\frac{16}{15\pi^2}m^{3/2}V \lbrack
nU''(n)\rbrack^{5/2}\;.\label{2.32} \ee We then find from
(\ref{2.13}) that \be
\frac{1}{V}\Omega_{T=0}^{(1)}=\frac{8\hbar}{15\pi^2}m^{3/2}\lbrack
nU''(n)\rbrack^{5/2}\;.\label{2.33} \ee

If we use (\ref{2.28a}) we have \be
\rho_{T\ne0}^{(1)}=\hbar\int\frac{d^3k}{(2\pi)^3}\frac{E_k}{
(e^{\beta E_k}-1)}\;,\label{2.34} \ee where $E_k$ is given by
(\ref{2.29}) after the large volume limit has been taken. Using
spherical polar coordinates, performing the angular integration,
and rescaling $k\rightarrow(2mT)^{1/2}k$ results in \bea
\rho_{T\ne0}^{(1)}&=&\frac{\hbar}{2\pi^2\beta}
\left(\frac{2m}{\beta}\right)^{3/2} \int_{0}^{\infty}dk\,k^3
\sqrt{k^2+2\beta nU''(n)}\nn &&\quad\quad\times\left\lbrack
e^{k\sqrt{k^2+2\beta nU''(n)}}-1\right\rbrack^{-1}\;.\label{2.35}
\eea Although it is not possible to evaluate the integral in
(\ref{2.35}) exactly, it is relatively straightforward to find the
low (or high) temperature expansion. We concentrate on the low
temperature expansion here and assume that $2\beta nU''(n)>>1$.
The first two terms in the asymptotic expansion of (\ref{2.35})
give \bea \rho_{T\ne0}^{(1)}&\simeq&\frac{\hbar\pi^2}{30}\left(
\frac{m}{nU''(n)}\right)^{3/2} T^4 \Big\lbrace1-
\frac{25\pi^2}{21}\lbrack\beta nU''(n)\rbrack^{-2}\nn
&&\quad\quad+\cdots\Big\rbrace \;.\label{2.36} \eea We then have
the low temperature expansion of the energy density in
(\ref{2.28}) being \bea
\rho(T)&\simeq&U(n)+\frac{8\hbar}{15\pi^2}m^{3/2}\lbrack nU''(n)\rbrack^{5/2}\label{2.37}\\
&&\hspace{-0.5cm}+\frac{\hbar\pi^2T^4}{30}\left\lbrack\frac{m}{nU''(n)}\right\rbrack^{3/2}
\left\lbrace1-\frac{25\pi^2}{21}\lbrack\beta nU''(n)\rbrack^{-2}\right\rbrace.\nonumber
\eea
In the zero temperature limit only the first two terms survive and we have the exact result
\be
\rho(T=0)=U(n)+\frac{8\hbar}{15\pi^2}m^{3/2}\lbrack nU''(n)\rbrack^{5/2} \;.\label{2.38}
\ee

So far we have kept the potential $U(\b2)$ arbitrary. In order  to
make contact with the standard results we will take \be
U(\b2)=\frac{2\pi a}{m}|\bp|^4\;,\label{2.39} \ee with $a$ the
$s$-wave scattering length. In this case it follows from
(\ref{2.38}) that \be \rho(T=0)=\frac{2\pi
an^2}{m}\left\lbrack1+\frac{128\hbar}{15\sqrt{\pi}}\sqrt{na^3}
\right\rbrack\;,\label{2.40} \ee and from (\ref{2.36}) that \be
\rho_{T=0}^{(1)}\simeq\frac{\hbar
m^3T^4}{240\sqrt{\pi}}(na^3)^{-3/2}\left\lbrace1-\frac{25}{336}
\left(\frac{mT}{na}\right)^2+\cdots\right\rbrace\;.\label{2.41}
\ee The result in (\ref{2.40}) agrees with the classic
calculations of Refs.~\cite{LeeYang,BruecknerSawada,LeeHuangYang}
for the zero temperature gas. The first term of (\ref{2.40})
reproduces the leading correction at low temperature found in
Refs.~\cite{LeeYangII}. (Note that Ref.~\cite{LeeYangII} chooses
$m=1/2$.)

\subsection{Equation of state}
\label{sec2.3}

The equation of state involves a determination of the pressure,
which follows from the first law of thermodynamics as \be
P=-\left.\left(\frac{\partial\Omega}{\partial
V}\right)\right|_{T,\mu,\bp} \;.\label{2.42} \ee Because $\Omega$
is an extensive quantity, in the large volume limit we have simply
\be P=-\frac{\Omega}{V}\;.\label{2.43} \ee The evaluation of the
equation of state involves the determination of $\Omega$ to order
$\hbar$. From (\ref{2.2}) and (\ref{2.3}) we have \be
P=\mu\b2-U(\b2)-\frac{\Omega_{T=0}^{(1)}+\Omega_{T\ne0}^{(1)}}{V}
\;.\label{2.44} \ee Using (\ref{2.20}) and (\ref{2.23}) to
eliminate $\b2$ we find (to order $\hbar$) \be P=\mu
n-U(n)-\frac{\Omega_{T=0}^{(1)}+\Omega_{T\ne0}^{(1)}}{V}
\label{2.45} \ee where we can set $\b2=n$ in $\Omega_{T=0}^{(1)}$
and $\Omega_{T\ne0}^{(1)}$ since these two terms are already of
order $\hbar$. What makes the calculation of $P$ more difficult
than that of $\rho$ is that we cannot simply use (\ref{2.18}) for
$\mu$ in the first term; we need to know the ${\mathcal O}(\hbar)$
contribution to $\mu$ coming from the quantum corrections to the
classical field equation for $\bp$.

Because we are interested only in the condensed phase, we can
replace (\ref{2.17}) with \be
0=\left.\left(\frac{\partial\Omega}{\partial\b2}\right)
\right|_{\beta,V,\mu}\;.\label{2.46} \ee This leads to \be
\mu=U'(\b2)+\left.\frac{1}{V}\frac{\partial( \Omega_{T=0}^{(1)}+
\Omega_{T\ne0}^{(1)})}{\partial\b2}\right|_{\mu,T,V}
\;.\label{2.47} \ee We now use (\ref{2.20}) again in the first
term to find \bea \mu&=&U'(n)+\frac{1}{V}\left\lbrace
\frac{\partial(
\Omega_{T=0}^{(1)}+\Omega_{T\ne0}^{(1)})}{\partial\b2}\right.\nn
&&\left.\quad\quad\quad+ U''(n)
\frac{\partial(\Omega_{T=0}^{(1)}+\Omega_{T\ne0}^{(1)})}{\partial\mu}
\right\rbrace\;.\label{2.48} \eea Using this in (\ref{2.45})
results in \bea P&=&nU'(n)-U(n)+ \frac{1}{V}\left\lbrace
n\frac{\partial \Omega_{T=0}^{(1)}}{\partial\b2}\right.\nn
&&\left.+nU''(n)\frac{\partial\Omega_{T=0}^{(1)}}{\partial\mu}-\Omega_{T=0}^{(1)}\right\rbrace+
\frac{1}{V}\left\lbrace n\frac{\partial
\Omega_{T\ne0}^{(1)}}{\partial\b2}\right.\nn
&&\left.+nU''(n)\frac{\partial\Omega_{T\ne0}^{(1)}}{\partial\mu}-
\Omega_{T\ne0}^{(1)}\right\rbrace\label{2.49} \eea for the
pressure to order $\hbar$. The last term of (\ref{2.49}) displays
the finite temperature correction to the zero temperature result.
In the last two terms of (\ref{2.49}) we can set $\b2=n$ and
$\mu=U'(n)$ to simplify the results after we have performed the
indicated differentiations.

We regularize the terms in $\Omega_{T=0}^{(1)}$ by the
$\zeta$-function method as described in Sec.~\ref{sec2.1}. From
the excitation energy (\ref{2.9}) it is easy to see that \be
\frac{\partial
E_k}{\partial\mu}=-\frac{1}{E_k}\left(\frac{\sigma_k}{2m}+nU''(n)\right)
\;\label{2.50} \ee \bea \frac{\partial E_k}{\partial\b2}&=&
\frac{1}{E_k}\Big\lbrace\Big(2U''(n)+nU'''(n)\Big)\frac{\sigma_k}{2m}
\nn &&\quad+n(U''(n))^2\Big\rbrace\;.\label{2.51} \eea It then
follows that \bea \frac{ n} {V}\left\lbrace \frac{ \partial
\Omega_{T = 0}^{(1)}} {\partial\b2} + U'' (n) \frac{
\partial\Omega_{T = 0}^{(1)}} {\partial\mu}\right\rbrace & = &
\frac{ \hbar n} {2V} (1-s) \ell^{-s}\nn &
&\hspace{-3cm}\times\lbrack U'' (n) + nU'''
(n)\rbrack\sum_kE_k^{-1-s} \frac{ \sigma _k} {2m}\label{2.52} \eea
with the right hand side analytically continued to $s = 0$.
Taking the large volume limit it is  easy to show that \bea \frac{
n} {V}\left\lbrace \frac{ \partial \Omega_{T = 0}^{(1)}}
{\partial\b2} + U'' (n) \frac{ \partial\Omega_{T = 0}^{(1)}}
{\partial\mu}\right\rbrace & = & \frac{ \hbar n}
{8\pi^2}(2m)^{3/2}\nn &&\hspace{-3cm}\times\lbrack U'' (n) + nU'''
(n)\rbrack\ell^{-s}(1-s)\nn
&&\hspace{-3cm}\times\frac{\Gamma(2-s/2)\Gamma(s-3/2)}{\Gamma\left(\frac{1+s}{2}\right)}
\lbrack2mU''(n)\rbrack^{3/2-s}\nn &&\hspace{-4cm}=\frac{4\hbar
n}{3\pi^2}\lbrack U'' (n) + nU''' (n)\rbrack\lbrack
mnU''(n)\rbrack^{3/2}\label{2.53} \eea after analytic continuation
to $s = 0$.

The pressure of the zero temperature gas follows from (\ref
{2.49}) as \bea P(T = 0)& = & nU' (n) -U(n) +\frac{ 4 \hbar} {5
\pi^2} m^{3/2}\lbrack nU'' (n)\rbrack^{5/2}\nn & &+ \frac{ 4
\hbar} {3 \pi^2} n^2U''' (n)\lbrack mnU''
(n)\rbrack^{3/2}\label{2.54} \eea if we use (\ref{2.33}) and
(\ref{2.53}). Specialising to $U (n)  = 2 \pi an^2/m$ as in
(\ref{2.39}) results in \be P_{T = 0} = \frac{ 2 \pi an ^ 2}
{m}\left\lbrack 1 + \frac{ 64 \hbar} {5\sqrt{\pi}}\sqrt{na ^
3}\right\rbrack\label{2.55} \ee in agreement with the standard
results\cite{LeeYang,LeeHuangYang}.

To evaluate the finite temperature part of the pressure, we use
(\ref{2.11}) along with (\ref{2.50}) and (\ref{2.51}) to find \bea
\frac{ n} {V}\left\lbrace \frac{ \partial  \Omega_{T \ne 0}^{(1)}}
{\partial\b2} + U'' (n) \frac{ \partial\Omega_{T \ne 0}^{(1)}}
{\partial\mu}\right\rbrace & = &\frac{\hbar n}{V}\lbrack U'' (n) +
nU''' (n)\rbrack\nn &&\hspace{-4.2cm}\times\sum_k(e^{\beta
E_k}-1)^{-1}\left\lbrack\frac{\sigma_k}{\sigma_k+4mnU''(n)}\right\rbrack^{1/2}\;.\label{2.56}
\eea In the large box limit we can evaluate the right hand side at
low temperatures as we did for $\rho _ {T\ne0} ^ {(1)} $ in
Sec.~\ref{sec2.2}.  A short calculation leads to \bea \frac{ n}
{V}\left\lbrace \frac{ \partial \Omega_{T \ne 0}^{(1)}}
{\partial\b2} + U'' (n) \frac{ \partial\Omega_{T \ne 0}^{(1)}}
{\partial\mu}\right\rbrace & = &\frac{\hbar
\pi^2T^4}{60}\left\lbrack \frac{m}{nU'' (n)}\right\rbrack^{3/2}\nn
&&\hspace{-5cm}\times\left\lbrack1+\frac{nU'''(n)}{U''(n)}
\right\rbrack\left\lbrace1-\frac{5\pi^2}{3}\lbrack\beta
nU''(n)\rbrack^{-2}+\cdots\right\rbrace\label{2.57} \eea where
$\beta nU''(n)>>1$. In a similar way the low temperature expansion
of (\ref{2.11}) gives \bea -\frac{\Omega_{T\ne0}^{(1)}}{V}&=
&\frac{\pi^2\hbar}{90}T^4\left\lbrack\frac{m}{nU''(n)}\right
\rbrack^{3/2}\Big\lbrace1\nn &&-\frac{5\pi^2}{7}\lbrack\beta
nU''(n)\rbrack^{-2}+\cdots\Big\rbrace\;.\label{2.58} \eea
Combining (\ref{2.57}) and (\ref{2.58}) we have as the finite
temperature correction to the pressure \bea
P_{T\ne0}&\simeq&\frac{\pi^2\hbar}{36}T^4\left\lbrack\frac{m}{nU''(n)}\right\rbrack^{3/2}
\left\lbrace1+\frac{3nU'''(n)}{5U''(n)}\right.\nn
&&\hspace{-3pt}\left.-\pi^2\left(\frac{9}{7}+\frac{nU'''(n)}{U''(n)}\right)\lbrack\beta
nU''(n)\rbrack^{-2}+\cdots\right\rbrace\label{2.59} \eea
 at low temperatures.  When we specialise to $U (n) = 2 \pi an ^ 2/m$ this result reduces to
\be P_{T\ne0}\simeq \hbar \frac{\sqrt {\pi}} {288} m ^ 3T ^ 4
(na)^{-3/2} \left\lbrace 1-\frac{ 9} {118}\left (\frac{ m} {\beta
na}\right)^2 +\cdots\right\rbrace\;.\label{2.60} \ee The leading
term agrees with the earlier calculation in Ref.~\cite{LeeYangII}.

\section{Binary mixture}
\label{sec3}

\subsection {General approach}
\label{sec3.1}

We now generalize the results of Sec.~\ref{sec2} to a  two-species
Bose gas described by Schr\"{o}dinger fields $\Psi_1$ and
$\Psi_2$.  We allow for self-interactions as well as mutual
interactions between the two species in a general way.  The action
functional will be chosen to be \bea S & = &\int\!\!
dt\!\!\int_\Sigma\!\! d\sigma_x\Big\lbrack\sum_{j =
1}^{2}\Big\lbrace \frac{ i} {2}\left(\Psi_j ^\dagger\dot {\Psi}
_j-\dot {\Psi} _ {j} ^\dagger\Psi_j\right)-\frac{ 1} {2m_j} \left
|\nabla\Psi_j\right|^2\nn & & +\mu_j\left
|\Psi_j\right|^2\Big\rbrace -U\left (|\Psi_1 | ^ 2, |\Psi_2 | ^
2\right)\Big\rbrack.\label{3.1} \eea Here $m_ {1,2} $ denote the
masses of the two species, and $\mu_ {1,2} $ are the chemical
potentials.  We assume that the interaction, described by the
potential $U$, only involves the magnitudes of the two fields.
The theory has a rigid $U (1)\times U (1) $ symmetry in general,
unless we further specify the potential.  The field equations
which follow from (\ref{3.1}) are \bea 0 &=&  i\dot {\Psi} _j +
\frac{ 1} {2m_j}\nabla^2\Psi_j +\mu_j\Psi_j\nn
&&\quad\quad-U_j\left (|\Psi_1 |^2, |\Psi_2 |
^2\right)\Psi_j\label{3.2} \eea for $j = 1,2$ along with the
hermitian conjugate.  Here $U_j = \frac{ \partial} {\partial
|\Psi_j |^2} U$.

The first requirement is to identify the excitation energies  by
first setting $\Psi_j =\bp_j +\Phi_j$ and linearizing in $\Phi_j$.
As in Sec.~\ref{sec2.1}, $\bp_j$ denotes the background field
which we assume to be constant.  We find \bea
0&=&i\dot{\Phi}_1+\frac{1}{2m_1}\nabla^2\Phi_1+\mu_1\Phi_1-(U_1+|\bp_1
|^2U_{11})\Phi_1\nn
&&-\bp_1^2U_{11}\Phi_1^\dagger-\bp_1\bp_2^\dagger
U_{12}\Phi_2-\bp_1\bp_2 U_{12}\Phi_2^\dagger,\label{3.3} \eea
 along with its hermitian conjugate, and two other equations
 obtained by interchanging the labels 1 and 2.  We have
 abbreviated $U_ {jj'} = \frac{ \partial ^ 2} {\partial
 |\Psi_ {j'} | ^ 2 \partial |\Psi_j | ^ 2} U = U_ {j' j} $.
 All terms involving the potential are evaluated at a background
 fields.  To get the excitation energies we set
\bea
\Phi_j (t,\x) & = &\sum _kA_ {kj} e ^ {-iE_kt} f_k (\x)\;,\label {3.4a}\\
\Phi_j ^\dagger(t,\x) & = &\sum _kB_ {kj} e ^ {-iE_kt} f_k (\x)\;,\label {3.4b}
\eea
for independent coefficients $A_ {kj} $ and $B_ {kj} $.  ($f_k (\x) $
is still the solution to (\ref {2.6}).)  After some calculation the
excitation energies can be shown to be
\bea
E_ {k\pm} ^ 2 & = & \frac{ 1} {2}\left (a_1 ^ 2 + a_2 ^ 2-|\bp_1
| ^ 4U_ {11} ^ 2-|\bp_2 | ^ 4U_ {22} ^ 2\right)\nn
& &\pm \frac{ 1} {2}\Big\lbrack\left (a_1 ^ 2-a_2 ^ 2-|\bp_1
| ^ 4U_ {11} ^ 2 + |\bp_2 | ^ 4U_ {22} ^ 2\right) ^ 2\nn
& &+ 16 |\bp_1 | ^ 2 |\bp_2 | ^ 2U_ {12} ^ 2\nn
&&\times (a_1-|\bp_1 | ^ 2U_ {11}) (a_2-|\bp_2 | ^ 2U_ {22})\Big\rbrack ^ {1/2}
\label{3.5}
\eea
where
\be
a_j = \frac{\sigma_k} {2m_j} -\mu_j + U_j + |\bp_j | ^ 2U_ {jj}\ \ (j = 1,2).\label {3.6}
\ee

The thermodynamic potential $\Omega$ has the form (\ref {2.1}) where
\bea
\Omega _ {\rm class} & = & V\left\lbrack U (|\bp_1 | ^ 2, |\bp_2 | ^ 2)
-\mu_1 |\bp_1 | ^ 2-\mu_2 |\bp_2 | ^ 2\right\rbrack,\label {3.7}\\
\Omega_ {T = 0} ^ {(1)} & = & \frac{ \hbar} {2}\sum _k (E_ {k +} + E_ {k-}),\label {3.8}\\
\Omega _ {T\ne0} ^ {(1)} & = & \hbar T\sum _k\left\lbrack\ln\left (1-e ^ {-\beta E_ {k +}}
\right)\right. \nn
&&\left.\hspace{2cm}+\ln\left (1-e ^ {-\beta E_ {k-}}\right)\right\rbrack.\label {3.9}
\eea
The number density for species $j$ is (see (\ref {2.14}))
\be
n_j = -\frac{ 1} {V}\left.\left (\frac{ \partial\Omega} {\partial\mu_j}
\right)\right|_ {T, V,\bp}\ \ (j = 1,2)\label {3.10}
\ee
and the energy density is given by (\ref {2.15}).  It is easy to show
that in place of (\ref {2.16}) we have
\be
\rho = \frac{ 1} {V} \frac{ \partial} {\partial \beta}\left.\left
(\beta\Omega\right)\right |_{\mu, V,\bp} +\mu_1n_1 +\mu_2n_2\;.\label {3.11}
\ee
The pressure is given by (\ref{2.43}).

\subsection {Ground state energy}
\label {sec3.2}

Using (\ref{3.10}) we can write
\be
n_j = |\bp_j | ^ 2 + n_ {jT = 0} ^ {(1)} + n_ {jT\ne0} ^ {(1)}\ \ (j = 1,2)\label {3.12}
\ee
as in (\ref{2.20},\ref {2.21}).   This allows us to conclude that
$|\bp_j | ^ 2 = n_j + {\mathcal O } (\hbar) $.  Since only
$\Omega_ {T\ne0} ^ {(1)} $ has an explicit $\beta $ dependence, we find
\bea
\rho & = & U (|\bp_1 | ^ 2, |\bp_2 | ^ 2) + \frac{ 1} {V}
\Omega _ {T = 0} ^ {(1)} + \frac{ 1} {V} \frac{ \partial}
{\partial \beta}\left.\left (\beta\Omega_{T\ne0} ^ {(1)}\right)\right|_{\mu, V,\bp}\nn
& & +\mu_1 (n_1-|\bp_1 | ^ 2) +\mu_2 (n_2-|\bp_2 | ^ 2).\label {3.13}
\eea
To order $\hbar $ we have $\mu_j = U_j$ as in (\ref{2.18}).
We can simplify (\ref{3.13}) to order $\hbar $ as
\be
\rho = U (n_1,n_2) + \frac{ 1} {V}\Omega_{T = 0} ^ {(1)} +
\rho _ {T\ne0}\label {3.14}
\ee
where
\bea
\rho_ {T\ne0} & = & \frac{ 1} {V} \frac{ \partial} {\partial \beta}
\left.\left (\beta\Omega_{T\ne0} ^ {(1)}\right)\right|_ {\mu, V,\bp}\nn
& = & \frac{ 1} {V}\sum_k\Big\lbrace E_{k +}\left (e ^ {\beta E_{k +}} -1\right) ^ {-1}\nn
&&\hspace{1cm} +E_{k -}\left (e ^ {\beta E_{k -}} -1\right) ^ {-1}\Big\rbrace\label {3.15}
\eea
is the finite temperature correction to the zero-temperature result.

  We were not able to obtain analytical results for the ground
  state energy density except in two special cases.  One case
  is $U_ {12} = 0$, which means that the two species do not
  interact with each other.  This is not very interesting since
  it just results in a sum of terms for the two species, with
  each term given as in Sec.~\ref {sec2}.  The second case
  where we can obtain exact analytic results at $T = 0$ occurs
  if we take $m_1 = m_2$.  A physical system where this is relevant
  concerns different hyperfine spin states of a given gas.  This
  specialisation allows a sufficient simplification of (\ref{3.5})
  to enable analytic results.

We are free to set $|\bp_j | ^ 2 = n_j$ in the second two terms of
(\ref{3.14}), as well as to take $\mu_j = U_j$.  With these two
simplifications made, and $m_1 = m_2 = m$ taken, the excitation
energies (\ref{3.5},\ref{3.6}) become
\be
E_ {k\pm} ^ 2 = \frac{ \sigma _k} {2m}\left (\frac{ \sigma _k} {2m} +\alpha _\pm\right)\label {3.16}
\ee
where
\bea
\alpha _\pm &=& n_1U_ {11} + n_2U_ {22}\nn
&&\pm\left\lbrack (n_1U_ {11} -n_2U_ {22}) ^ 2 + 4n_1n_2U_ {12} ^ 2\right\rbrack ^ {1/2}.\label {3.17}
\eea

The result for $\Omega _ {T = 0} ^ {(1)} $ may be defined using the
energy $\zeta $-function method exactly as in Sec.~\ref {sec2}.  In
fact the energy $\zeta $-function for the energies $E_ {k\pm} $ may
be obtained from (\ref{2.31}) by replacing $2nU'' $ with $\alpha _\pm$.
The energy density of the zero temperature gas is therefore
\be
\rho _ {T = 0} = U (n_1,n_2) + \frac{ \hbar} {30 \pi ^ 2}
(2m) ^ {3/2}\left (\alpha _ + ^ {5/2} +\alpha _-^ {5/2}\right)\;.\label {3.18}
\ee

The finite temperature correction, at low temperatures, can be
obtained from (\ref{2.36}) with the replacement $2nU'' $ with
$\alpha _\pm$. The leading term at low temperature is \be \rho _
{T\ne0}\simeq \hbar \frac{ \pi ^ 2} {30} T ^ 4\left\lbrack\left
(\frac{ 2m} {\alpha _ +}\right) ^ {3/2} +\left (\frac{ 2m} {\alpha
_-}\right) ^ {3/2}\right\rbrack\;.\label {3.19} \ee If we set $U_
{12} = 0$, then the energy density reduces to the sum of the
single species gas contributions.  The presence of mutual
interactions of the two species complicates the result.

It is possible to generalize the method described for the single
species gas in Sec.~\ref{sec2.3} to deal with the binary mixture.
However as the details are rather messy, and the result not
particulcarly simple we will not pursue this here.

\section{Discussion and conclusions}
\label{sec4}

We have shown how $\zeta $-function techniques may be used to
obtain the thermodynamic properties of a single species as well as
a two species Bose gas.  The interaction in either case was
arbitrary.  For the binary mixture analytical results at zero
temperature were obtained in the case where the masses of the two
constituents were equal.  Although this restriction might seem
artificial at first sight, and perhaps not of great physical
relevance, it has been realised experimentally for trapped
gases\cite{Myatt}.  Some progress can be made in the more general
case of unequal masses, but analytical results do not appear
possible.

The overwhelming interest at the present time is in the behaviour
of trapped Bose gases.  In the present formalism this entails the
introduction of an additional term to (\ref{2.1}) to represent the
trapping potential.  This loses the feature of the calculation
above which enabled analytical results to be obtained; namely,
that $\bar{\Psi}$ was constant.  Nevertheless the background field
method works whether $\bar{\Psi}$ is constant or not, and it seems
likely that the use of generalised $\zeta $-function methods like
those presented above can be used to provide a different approach
to the problem of trapped interacting gases.

\end{document}